\documentclass{article}
\usepackage{amssymb}
\usepackage{amsfonts}
\usepackage{amsmath}

\input{tcilatex}

\begin{document}

\title{Quantum particles trapped in a position-dependent mass barrier; a
d-dimensional recipe}
\author{Omar Mustafa$^{1}$ and S. Habib Mazharimousavi$^{2}$ \\
Department of Physics, Eastern Mediterranean University, \\
G Magusa, North Cyprus, Mersin 10,Turkey\\
$^{1}$e-mail: omar.mustafa@emu.edu.tr\\
$^{2}$e-mail: habib.mazhari@emu.edu.tr}
\maketitle

\begin{abstract}
We consider a free particle, $V\left( r\right) =0,$ with a
position-dependent mass $m(r)=1/(1+\varsigma ^{2}r^{2})^{2}$ in the
d-dimensional Schr\"{o}dinger equation. The effective potential turns out to
be a generalized P\"{o}schl-Teller potential that admits exact solution.

\end{abstract}

\section{Introduction}

Quantum mechanical particles endowed with position - dependent mass (PDM), $%
M\left( r\right) $, have attracted attention and inspired intense research
activities over the years [1-16] They form interesting and useful models for
many physical problems. They are used, for example, in the energy density
many-body problem [2], in the determination of the electronic properties of
the semiconductors [3] and quantum dots [4], in quantum liquids [5], in the $%
^{3}He$ clusters [6] and metal clusters [7], in the Bohmian approach to
quantum theory [8], full/partial Gaussian wave-packet revival inside an
infinite potential [9], etc.

In addition to its practical applicability side, conceptual problems of
delicate nature erupt in the study of quantum mechanical systems with
position-dependent mass (e.g., momentum operator does not commute with $%
m\left( r\right) $, uniqueness of kinetic energy operator, etc.).
Comprehensive discussion on such issues could be found in, e.g., [8,10, and
related references therein].

On the other hand, the exact solvability of the $d$-dimensional Schr\"{o}%
dinger equation may very well form the major ingredient for methods that are
based on a Liouvillean-type change of variables (cf., e.g., [17-19]) such as
the point canonical transformation (PCT) method\ (cf., e.g., [13,14]).
Where, in such methodologies, the exact solution of the so-called \emph{%
reference Schr\"{o}dinger equation} \ (eigenvalues and eigenfunctions) is
mapped into an exact solution (eigenvalues and eigenfunctions) of the
so-called \emph{target Schr\"{o}dinger equation\ }(cf, e.g. [1,14]).

In a recent study, we have introduced a $d$-dimensional regularization of
the PCT-method for some PDM-quantum mechanical particles [1]. Therein,
inter-dimensional degeneracies associated with the isomorphism between the
angular momentum quantum number $\ell $ and dimensionality $d$ are
incorporated through the central repulsive/attractive core $\ell \left( \ell
+1\right) /r^{2}\longrightarrow \ell _{d}\left( \ell _{d}+1\right) /r^{2}$
(where $\ell _{d}=\ell +\left( d-3\right) /2$ for $d\geq 2$) of the
spherically symmetric Schr\"{o}dinger equation (cf, e.g., [20], Gang in [10]
and Quesne in [10] for more details).

In this work, we study "\emph{free"} \ particles, $V\left( r\right) =0$,
trapped in their own position-dependent mass barriers (hence, labeled as 
\emph{quasi-free} particles), where inter-dimensional degeneracies remain
intact in the $d$-\emph{dimensional} \emph{radial Schr\"{o}dinger equation}
(in atomic units $\hbar =m_{\circ }=1$, the position-dependent mass $M\left(
r\right) =m_{\circ }\,m\left( r\right) ,$ and with $\alpha =\gamma =0$ and $%
\beta =-1$ in Eq.(1.1) of Tanaka in [10].) 
\begin{equation}
\left\{ \frac{d^{2}}{dr^{2}}-\frac{\ell _{d}\left( \ell _{d}+1\right) }{r^{2}%
}+\frac{m^{\prime }\left( r\right) }{m\left( r\right) }\left( \frac{d-1}{2r}-%
\frac{d}{dr}\right) +2m\left( r\right) E\right\} R_{n_{r},\ell }\left(
r\right) =0.
\end{equation}%
Where $n_{r}=0,1,2,\cdots $ is the radial quantum number, and $m^{\prime
}\left( r\right) =dm\left( r\right) /dr.$ Moreover, the $d=1$ can be
obtained through $\ell _{d}=-1$ and $\ell _{d}=0$ \thinspace for even and
odd parity, $\mathcal{P=}\left( -1\right) ^{\ell _{d}+1}$, respectively (cf,
e.g., Mustafa and Znojil in [20]). Nevertheless, the inter-dimensional
degeneracies associated with the isomorphism between angular momentum $\ell $
and dimensionality $d$ \ builds up the ladder of excited states for any
given $n_{r}$ and nonzero $\ell $ from the $\ell =0$ result, with that $%
n_{r} $, by the transcription $d\rightarrow d+2\ell .$ That is, if $%
E_{n_{r},\ell }\left( d\right) $ is the eigenvalue in $d$-dimensions then%
\begin{equation}
E_{n_{r},\ell }\left( 2\right) \equiv E_{n_{r},\ell -1}\left( 4\right)
\equiv \cdots \equiv E_{n_{r},1}\left( 2\ell \right) \equiv
E_{n_{r},0}\left( 2\ell +2\right)
\end{equation}%
for even $d$, and%
\begin{equation}
E_{n_{r},\ell }\left( 3\right) \equiv E_{n_{r},\ell -1}\left( 5\right)
\equiv \cdots \equiv E_{n_{r},1}\left( 2\ell +1\right) \equiv
E_{n_{r},0}\left( 2\ell +3\right)
\end{equation}%
for odd $d$. For more details on inter-dimensional degeneracies the reader
may refer to a sample of references in [20].

With the PCT-method in point (cf, e.g., [1]), a substitution of the form $%
R\left( r\right) =g\left( r\right) \,\phi \left( q\left( r\right) \right) $
in (1) would result in $g\left( r\right) ^{2}q^{\prime }\left( r\right)
=m\left( r\right) $, manifested by the requirement of a vanishing
coefficient of the first-order derivative of $\phi \left( q\left( r\right)
\right) $ ( hence a one-dimensional form of Schr\"{o}dinger equation is
achieved), and $q^{\prime }\left( r\right) ^{2}=m\left( r\right) $ to avoid
position-dependent energies-multiplicity (i.e., $2E\,m\left( r\right)
/q^{\prime }\left( r\right) ^{2}\Longrightarrow 2E$). Hence, 
\begin{equation}
q\left( r\right) =\int^{r}\sqrt{m\left( t\right) }dt\text{ }\implies g\left(
r\right) =m\left( r\right) ^{1/4}.
\end{equation}%
This in effect implies%
\begin{equation}
\left\{ -\frac{1}{2}\frac{d^{2}}{dq^{2}}+V_{eff}\left( q\left( r\right)
\right) \right\} \phi _{n_{r},\ell _{d}}\left( q\left( r\right) \right)
=E_{d}\phi _{n_{r},\ell _{d}}\left( q\left( r\right) \right) ,
\end{equation}%
with an effective potentials%
\begin{equation}
V_{eff}\left( q\left( r\right) \right) =\frac{\ell _{d}\left( \ell
_{d}+1\right) }{2r^{2}m\left( r\right) }-U_{d}\left( r\right)
\end{equation}%
where%
\begin{equation}
U_{d}\left( r\right) =\frac{m^{\prime \prime }\left( r\right) }{8m\left(
r\right) ^{2}}-\frac{7m^{\prime }\left( r\right) ^{2}}{32m\left( r\right)
^{3}}+\frac{m^{\prime }\left( r\right) \left( d-1\right) }{4r\,m\left(
r\right) ^{2}}.
\end{equation}

\section{Consequences of an asymptotically vanishing mass settings as $%
r\longrightarrow \infty $}

A "\emph{free" }particle with an asymptotically vanishing position-dependent
mass\emph{\ }$m(r)=1/(1+\varsigma ^{2}r^{2})^{2}$ would experience an
effective potential%
\begin{equation}
V_{eff}\left( q\left( r\right) \right) =\frac{\varsigma ^{2}}{2}\left[ \frac{%
\varkappa (\varkappa -1)}{\sin ^{2}(\varsigma q)}+\frac{\lambda (\lambda -1)%
}{\cos ^{2}(\varsigma q)}\right] -\frac{\varsigma ^{2}}{2},
\end{equation}%
where%
\begin{equation}
q(r)=\frac{1}{\varsigma }\arctan (\varsigma r)\Rightarrow \varsigma r=\tan
(\varsigma q),
\end{equation}%
and%
\begin{equation}
U_{d}(r)=-\left( \varsigma ^{2}d\right) \tan ^{2}\left( \varsigma q\right) +%
\frac{\varsigma ^{2}}{2}(1-2d)
\end{equation}%
In such settings, Eq. (5) reads%
\begin{equation}
\left\{ -\frac{1}{2}\frac{d^{2}}{dq^{2}}+\frac{\varsigma ^{2}}{2}\left[ 
\frac{\varkappa (\varkappa -1)}{\sin ^{2}(\varsigma q)}+\frac{\lambda
(\lambda -1)}{\cos ^{2}(\varsigma q)}\right] \right\} \phi _{n_{r},\ell
_{d}}\left( q\right) =\varepsilon \phi _{n_{r},\ell _{d}}\left( q\right) ,
\end{equation}%
where 
\begin{equation}
\varkappa (\varkappa -1)=l_{d}(l_{d}+1),\text{ }\lambda (\lambda
-1)=l_{d}(l_{d}+1)+2d\text{ \ \ and }\varepsilon =E+\frac{1}{2}\varsigma
^{2}.
\end{equation}%
Equation (11) is obviously a standard one-dimensional form of Schr\"{o}%
dinger equation with a generalized P\"{o}schl-Teller effective potential
which admits exact solution of the form 
\begin{equation}
\epsilon _{n_{r}}=\frac{\varsigma ^{2}}{2}(\varkappa +\lambda +2n_{r})^{2}
\end{equation}%
\begin{equation}
\phi _{n_{r},\ell _{d}}\left( q\right) =C\sin ^{\varkappa }(\varsigma q)\cos
^{\lambda }(\varsigma q)\,\,_{2}F_{1}(-n_{r},\varkappa +\lambda
+n_{r},\varkappa +\frac{1}{2};\sin ^{2}(\varsigma q))
\end{equation}%
with $\varkappa ,\lambda >1,$ $\phi _{n_{r},\ell _{d}}\left( 0\right) =0$
and $\phi _{n_{r},\ell _{d}}\left( \frac{\pi }{2\varsigma }\right) =0,$ as
reported by Salem and Montemayor (see Eq.(4.7) in [21]). This in turn would
lead to%
\begin{equation}
E_{n_{r},l_{d}}=\frac{\varsigma ^{2}}{2}((c+\frac{1}{2}\Delta
+2n_{r})^{2}-1);\,\text{\ }\Delta =\sqrt{(2l_{d}+1)^{2}+8d}
\end{equation}%
\begin{equation}
R_{n_{r},l_{d}}(r)=\tilde{C}\,\rho ^{l_{d}+1}(1+\rho ^{2})^{-\frac{1}{4}%
(2l_{d}+5+\Delta )}\text{ }\,_{2}F_{1}(-n_{r},c+\frac{\Delta }{2}+n_{r},c;%
\frac{\rho ^{2}}{1+\rho ^{2}})
\end{equation}%
where $\rho =\varsigma r,$ and $c=l_{d}+\frac{3}{2}$.

However, for $\varkappa =0,1$ (a requirement suggested by relation (12) when 
$\ell _{d}=0,-1$) the effective potential in (11) collapses into%
\begin{equation}
V_{eff}(q\left( r\right) )=\frac{\varsigma ^{2}}{2}\frac{\lambda (\lambda -1)%
}{\cos ^{2}(\varsigma q)}.
\end{equation}%
Which admits an exact solution%
\begin{equation}
E_{n_{r}}=2\varsigma ^{2}(n_{r}+\frac{\lambda }{2})^{2}-\frac{\varsigma ^{2}%
}{2}
\end{equation}%
\begin{equation}
\phi _{n_{r},\ell _{d}}(q)=A\text{ }\cos ^{\lambda }(\varsigma q)\text{ }%
_{2}F_{1}(-n_{r},n_{r}+\lambda ,\frac{1}{2},\sin ^{2}(\varsigma q)),
\end{equation}%
and consequently%
\begin{equation}
E_{n_{r},0}=2\varsigma ^{2}(n_{r}+\frac{\lambda }{2})^{2}-\frac{\varsigma
^{2}}{2}
\end{equation}%
\begin{equation}
R_{n_{r}}(r)=\tilde{A}\text{ }(1+\rho ^{2})^{-\frac{1}{4}(2l_{d}+5+\Delta )}%
\text{ }\,_{2}F_{1}(-n_{r},c+\frac{\Delta }{2}+n_{r},c;\frac{\rho ^{2}}{%
1+\rho ^{2}}).
\end{equation}%
where $\lambda =\left( 1+\Delta \right) /2.$

\section{Concluding Remarks}

In this letter, we considered a quasi-free particle with an asymptotically
vanishing position-dependent mass\emph{\ }$m(r)=1/(1+\varsigma
^{2}r^{2})^{2} $ and radial potential $V\left( r\right) =0$ (i.e., "free"
particle in this sense). We have shown that under these settings the
particle experiences an effective potential of the form of a P\"{o}%
schl-Teller, Eq.(8). The exact solution of which is mapped to match the
attendant settings of our quasi-free particle with the above mentioned
position-dependent mass.


\newpage

\begin{thebibliography}{99}
\bibitem{} O Mustafa and S.H Mazharimousavi 2006 "Point canonical
transformation d-dimensional regularization" (arXiv: math-ph/0602044)

\bibitem{} A Puente and M Casas Comput. Mater Sci. \textbf{2} (1994) 441

\bibitem{} Bastard G 1988 \emph{"Wave Mechanics Applied to Semiconductor
Heterostructures" ,} Les Editions de Physique, Les Ulis

\bibitem{} L I Serra and E Lipparini Europhys. Lett. \textbf{40} (1997) 667

\bibitem{} F Arias de Saaverda, J Boronat, A\ Polls , and A\ Fabrocini Phys.
Rev. \textbf{B 50} (1994) 4248

\bibitem{} M Barranco , M Pi , S.M Gatica ., E.S Hemandez ., and J. Navarro
Phys. Rev. \textbf{B 56} (1997) 8997

\bibitem{} A Puente , L I Serra, and M Casas Z. Phys. \textbf{D 31} ( 1994)
283

\bibitem{} A R Plastino, M Casas and A. Plastino Phys. Lett. \textbf{A281(}%
2001) 297 (and related references therein)

\bibitem{} A Schmidt Phys. Lett. \textbf{A} (2006) ( in press)

\bibitem{} T Tanaka J. Phys. \textbf{A 39} (2006) 219

C Quesne, Ann. Phys. \textbf{321} (2006) 1221

C Gang, Phys. Lett. A \textbf{329} (2004) 22

A R Plastino, A Rigo, M Casas, F Garcias, and A. Plastino Phys. Rev. \textbf{%
A 60 }(1999) 4318

\bibitem{} S H Dong and M. Lozada-Cassou Phys. Lett. \textbf{A 337} (2005)
313

I O Vakarchuk J. Phys. \textbf{A}; Math and Gen \textbf{38} (2005) 4727

C Y Cai, Z Z Ren and G X Ju Commun. Theor. Phys. \textbf{43} (2005)1019

\bibitem{} B Bagchi, A Banerjee, C Quesne and V M Tkachuk J. \ Phys. \textbf{%
A}; Math and Gen \textbf{38} (2005) 2929

J Yu and S H Dong Phys. Lett. \textbf{A 325} (2004)194

L Dekar, L Chetouani and T F Hammann J. Math. Phys. \textbf{39} (1998) 2551

\bibitem{} C Quesne and V M Tkachuk J. \ Phys. \textbf{A}; Math and Gen 
\textbf{37} (2004) 4267

L Jiang, L Z Yi, and C S Jia Phys. Lett. \textbf{A 345} (2005) 279

A D Alhaidari Int. J. Theor. Phys. \textbf{42} (2003) 2999

\bibitem{} A D Alhaidari Phys. Rev. \textbf{A 66} (2002) 042116

\bibitem{} R De, R Dutt and U Sukhatme J. \ Phys. \textbf{A}; Math and Gen 
\textbf{25 }(1992) L843

\bibitem{} G Junker J. \ Phys. \textbf{A}; Math and Gen \textbf{23} (1990)
L881

\bibitem{} J Liouville J. Math. Pure Appl. \textbf{1} (1837)16

\bibitem{} M Znojil and G L\'{e}vai J. Math. Phys. \textbf{42} (2001)1996

M Znojil "$\mathcal{PT}$-symmetric form of the Hulth\'{e}n potential" (2000)
(arXiv: math-ph/0002017)

\bibitem{} M Znojil and G L\'{e}vai Phys. Lett. \textbf{A 271} (2000) 327

\bibitem{} D R Herschbach et al \emph{"Dimensional Scaling in Chemical
Physics" }( Kluwer Academic Publishers 1993, Dordrecht, Netherlands.)

D R Herschbach J. Chem. Phys. \textbf{84} (1986) 838

H Taseli J. Math. Chem. \textbf{20} (1996) 235

O Mustafa and M Znojil J. Phys.A; Math and Gen. \textbf{35} (2002) 8929

O Mustafa \thinspace and M Odeh J. Phys. A; Math and Gen. \textbf{32} (1999)
6653

O Mustafa and M Odeh J. Phys. A; Math and Gen. \textbf{33} (2000) 5207

M M Nieto Am. J. Phys. \textbf{47} (1979) 1067

\bibitem{} L D Salem and R Montemayor Phys. Rev. \textbf{A 47} (1993) 105

S Fl\"{u}gge 1974, \emph{Practical Quantum Mechanics,} Springer, Berlin.
\end{thebibliography}
\end{document}